\documentclass[preprint,showpacs,floats,letterpaper,floatfix,,groupedaddress,eqsecnum]{revtex4}
\usepackage{subfigure}

\usepackage{amssymb,amsmath}
\usepackage[dvips]{graphicx}

\usepackage{dcolumn,epsfig}

\begin{document}

\title{Inertial frictional ratchets and their load bearing efficiencies}
\author{D. Kharkongor$^1$$^,$$^2$, W.L. Reenbohn$^3$ and Mangal C. Mahato$^1$$^,$}
\email{mangal@nehu.ac.in}
\affiliation{$^1$Department of Physics, North-Eastern Hill University,
Shillong-793022, India}
\affiliation{$^2$Department of Physics, St. Anthony's College,
Shillong-793001, India}
\affiliation{$^3$Department of Physics, National Institute of Technology 
Meghalaya, Shillong-793003, India}

\begin{abstract}

We investigate the performance of an inertial frictional ratchet in a 
sinusoidal potential driven by a sinusoidal external field. The dependence of 
the performance on the parameters of the sinusoidally varying friction, such 
as the mean friction coefficient and its phase difference with the potential, 
is studied in detail. Interestingly, under certain circumstances,
the thermodynamic efficiency of the ratchet against an applied load shows a non-monotonic
behaviour as a function of the mean friction coefficient. Also, in the large
friction ranges, the efficiency is shown to increase with increasing applied
load even though the corresponding ratchet current decreases as the applied
load increases. These counterintuitive numerical results are explained in the text.
\end{abstract}

\vspace{0.5cm}
\date{\today}

\maketitle

\section{Introduction}

The present spurt in the research activity on thermal ratchets began with the
experimental observation of asymmetric motion of protein molecules at a 
constant temperature along microtubules\cite{Svoboda}. The idea 
of ratchets did, however, exist since long. The most popular among 
them being the thought experiment on ratchet and pawl by 
Feynman\cite{Feynman}. Of course, in this case the ratchet and pawl 
are kept at two different temperatures. Immediately after the biological
experiment two important physical models were put forth: one by Prost and 
coworkers and named as flashing ratchet\cite{Prost} and the other by Magnasco 
called the rocked ratchet\cite{Magnasco}. Both these pioneering models use 
asymmetric periodic potentials and the particles are made to move in presence 
of thermal noise (nonzero temperature). In the case of flashing ratchets the 
potential is switched {\it{on}} and {\it{off}} for certain durations 
repeatedly so that during the off condition the particles diffuse 
symmetrically about the initial position of minimum of the initial well of the 
potential and during the on situation the particles roll down to the minimum 
of the present well of the potential. By suitably adjusting the durations of 
the off and on conditions depending on the temperature one achieves net 
(ratchet) current because the maxima of the potential on the left and right 
sides of a minimum happen to be located at different distances. On the other 
hand, in the case of rocked ratchet the system is subjected to a sinusoidal 
external forcing (null mean impulse per period). Since the potential is 
asymmetric the particles encounter different potential barriers when the 
forcings have different directions during a cycle leading to a net current. 
Both these models had overdamped initiations. A comprehensive review on 
ratchets is given by Reimann\cite{Reimann}. The scope of ratchet models has
since been extended much further\cite{Hanggi,Bressloff}. 

The rocked ratchet model, with asymmetric periodic potential, was extended to 
underdamped systems for the first time by Jung and coworkers\cite{Jung}. This
pioneering work extended the possibility of ratchet effect in underdamped
deterministic systems. Subsequently, the underdamped deterministic ratchet 
model has been investigated further\cite{Mateos,Saikia1,Kenfack,Dandogbessi}.
There has since then been many extensions to this underdamped model and 
applied to many noise induced systems including Josephson 
junctions\cite{Jakub,Spie1,Spie2}, the diffusion of adatoms on the 
crystal surface \cite{Barromeo}, dislocation of defects in metals\cite{Braun} 
and to other important situations wherein instead of inactive particles 
self-propelling particles were considered\cite{Ghosh,Ai}. Later on the ratchet 
current was shown to be possible in symmetric periodic potentials as well using 
periodic but asymmetric drives again with null mean imposed impulse per 
period\cite{Millonas,Flach,Hennig}. The asymmetric drive could be 
obtained in various ways. For example, a 'square-wave' drive with a large 
constant force for a small duration and a small constant force in the opposite 
direction for a longer duration\cite{Statmech} gives an appropriate asymmetric 
drive. A biharmonic drive being an another example\cite{Martinez,Wanda1} of 
asymmetric drive used to obtain ratchet current in overdamped as well as 
underdamped symmetric periodic potential systems. In most of the cases, in order  
to obtain ratchet current the medium (substrate) is considered spatially 
uniform so that the friction coefficient is constant in space. With uniform 
medium, as discussed earlier, ratchet current is obtained either in an 
asymmetric periodic potential and rocked periodically in time or in a symmetric
peridic potential but rocked peridically with a time asymmetric force of zero
mean over a period. However, there is an another important case where the
medium is considered to offer nonuniform diffusion coefficient for the motion
of the particle to obtain net asymmetric current.  

In an important work, B\"{u}ttiker\cite{Buttiker} had shown that the effect of 
a periodically varying diffusion coefficient on a periodic potential of same 
period but with a phase difference is equivalent to making the periodic 
potential tilt with a constant mean slope. However, diffusion coefficient  
results from a combination of two quantities: temperature and friction 
coefficient of the medium. Thus either the temperature or the friction 
coefficient (or both) can be varied to obtain the desired diffusion 
coefficient. Nonuniformity of temperature automatically puts the system in 
nonequilibrium situation and as argued by Landauer\cite{Landauer} there will 
be asymmetric particle crossings over potential barriers even without the 
application of external drive. This result was also exploited to obtain 
ratchet current\cite{Blanter,Benjamin}. Luchsinger\cite{Luchsinger}, considered
a similar system as B\"{u}ttiker's model but instead of a space-dependent 
temperature field, the space-dependent diffusion was induced by a 
position-dependent mobility. In this overdamped case, a net current was 
attained by considering a two-state model with different transition rates in 
order that detailed balance may be broken. Friction coefficient, however, is 
ineffective in static equilibrium situations and it comes into play only when 
the particles are in motion. It is a weak agent and, unlike non-uniform 
temperature, needs simultaneous application of external drive to obtain 
ratchet current. Nevertheless, the nonuniform friction breaks the left-right symmetry if varied 
periodically as the potential but with a phase difference. In the present work
we consider the friction coefficient to be periodic in space as the potential
but with a phase difference.  

A particle moving in a medium and experiencing space dependent friction 
depending on the inhomogeneity of the medium is not so unusual in nature.
Wahnstr\"{o}m, using a realistic Lennard-Jones interaction potential as a
practical illustration, showed that the space dependence of friction appears
because of coupling of adatom-motion degrees of freedom with the ionic
vibrations on the surface of substrates\cite{Wahn}. Nonuniform friction can
occur also due to nonuniform density of the medium through which the particle
moves. For example, a periodic nonuniform density in air can be established
by setting up a stationary sound wave just as in a Kundt's tube experiment.
In addition if electric charges are placed periodically in a line such that
the position of the charges do not coincide exactly with the projected nodes 
or antinodes of the stationary wave, then a charged particle moving along the
line of nodes will experience not only a periodic potential but also a
periodic variation of friction with somewhat shifted in phase. We thus have a 
model system that can be realized experimentally.

Ratchet effect has been obtained in symmetrically periodic 
overdamped\cite{ratchets,Bao1,Raishma,Gehlen} as well as 
underdamped\cite{Pramana1,Statmech,Saikia1,Wanda1,Donrich,Saikia2} 
frictionally nonuniform systems. In Ref.\cite{Gehlen} the authors consider the 
motion of a dimer consisting of two components connected by a harmonic spring 
in the same periodic potential and external force environment but experiencing 
different friction coefficient. In Ref.\cite{Pramana1,Statmech,Saikia1,Saikia2}
the external periodic drive frequency were taken of the order of the mean rate
of passages of the particle across the potential barriers. In the present work,
for reason to be made clear later, we use larger frequency of period drive as
in Ref.\cite{Wanda1,Donrich}. We consider the motion of underdamped particles 
in a sinusoidal potential $V(x)=-\sin x$ and driven by a sinusoidal external 
field $F(t)=F_0\cos \omega t$ with $F_0<<F_c$, where $F_c$ is the critical 
field amplitude at which the barrier of the combined effective potential 
$U(x)=V(x)-xF(t)$ just disappears. Since $F_0<<F_c$, the particles always 
encounter a finite potential barrier. It is the presence of thermal 
fluctuations that help the particles to overcome the potential barriers and
nonzero phase difference $\theta$ in the periodic friction coefficient
$\gamma(x)=\gamma_0(1-\lambda\sin(x+\theta))$ leads to asymmetry in the barrier
crossings to the left and right directions giving ratchet current. Here $\lambda$ is
the inhomogeneity parameter with $0\leq\lambda\leq 1$ and $\gamma_0$ is the mean value 
of the friction coefficient $\gamma_0$ over one period.

Since the frictional inhomogeneity is the only symmetry breaking agent 
considered, with $\theta\neq~n\pi,~n=0,\pm1,\pm2,~\cdots$, the underdamped 
ratchet yields only small currents and consequently has very small 
thermodynamic efficiency (defined as the ratio of work done by the system 
against an applied load to the total energy spent on the system). Admittedly, the 
efficiency of the frictional ratchet considered in the present case is 
negligible compared to the maximal efficiency (mostly, Stokes efficiency) 
($>$0.6) obtained by Spiechowicz and coworkers\cite{Spie1,Spie2}. Interestingly, 
Ref.\cite{Spie1} shows that a symmetric periodic potential system driven by a 
temporally symmetric periodic force exhibits more efficiency when an asymmetric
Poissonian white noise of mean $<\eta(t)>=F$ is applied than when subjected to a constant
tilt $F$. The present work, however, is not about obtaining large ratchet currents or 
more efficient current rectification but how the thermodynamic efficiency 
changes as an externally applied load is changed for various $\gamma_0$ and 
$\omega$ values. Interestingly, it shows, for instance, under certain 
circumstances, some counter-intuitive result that the efficiency increases as 
an externally applied load is increased. (The numerical results will be 
presented in detail in a section in the following.) It is to be noted that 
overdamped frictional ratchets yield relatively much larger 
currents\cite{ratchets}. However, the overdamped ratchets, operating in a 
different domain of parameter space, do not show the kind of interesting 
results that we obtain in the present underdamped case.

In the present work, we explore the role of friction coefficient $\gamma$ on the
ratchet current and the thermodynamic efficiency of the frictional ratchet, though small in magnitude, 
sometimes shows seemingly counterintuitive 
results. For the purpose of obtaining the thermodynamic efficiency of the 
frictional ratchet, in addition to the sinusoidal forcing, we apply an external
constant force (load) opposing the ratchet current. Naturally, the ratchet can 
sustain only a small load before the ratchet current vanishes and gives way to 
the constant force induced current in the direction of the load. This critical 
value of load increases with increasing value of the mean friction coefficient 
$\gamma_0$. In a previous work, unlike the symmetric potential used in the 
present case, currents were calculated in a sinusoidally driven asymmetric 
periodic potential by applying an additional constant load\cite{Bao}. In that 
work, for small asymmetry parameter $\Delta\leq 0.2$ of the potential, the 
applied load direction is such as to help increase the current. The purpose of 
the present work, however, as mentioned earlier, is to let the ratchet perform 
work against the load (current reducing load) and thereby investigate 
the thermodynamic efficiency of the ratchet.

In the next section II, the model of the system will be described. In section
III we give a brief discussion on why spatially periodic variation of friction 
coefficient of the medium should yield ratchet current and also summarize some 
results of an earlier work which has direct bearing on the present work. The 
numerical results based on our present investigations of the behavior of
ratchet current and mean absorbed energy without and with load applied to the
system will be presented in detail in Sec. IV. In the last section V our main 
results will be discussed and summarised.

\section{The Model}

We consider the motion of an ensemble of under-damped non-interacting Brownian
particles each of mass $m$ in a periodic potential $V(x)=-V_0\sin(kx)$ with 
$V_0$ as the amplitude of the potential and $k$ is its wave number. The
medium in which the particle moves is taken to be inhomogeneous in the sense
that it offers a spatially varying friction with coefficient
\begin{equation}
\gamma(x)=\gamma_0(1-\lambda\sin(kx+\theta))
\end{equation}
that leads the potential by a phase difference $\theta$.
Here, $\lambda$ is the inhomogeneity parameter, with $0\leq\lambda\leq 1$
and hence $\gamma_0(1-\lambda)\leq\gamma(x)\leq\gamma_0(1+\lambda)$.

In addition, the potential is rocked by a sub-threshold periodic time-dependent
forcing $F(t) = F_0\cos(\omega t)$, with $\omega$ = 2$\pi$/$\tau$ as the
rocking frequency and $\tau$ as the rocking period and $F_0$ as the forcing amplitude. 
The equation of motion of
the particle subjected to a thermal Gaussian white noise $\xi(t)$ at
temperature $T$ is given by the Langevin equation\cite{Sancho,Pramana},

\begin{equation}
m\frac{d^{2}x}{dt^{2}}=-\gamma(x)\frac{dx}{dt}-\frac{\partial{V(x)}}{\partial
x}+F(t)+\sqrt{\gamma(x) T}\xi(t).
\end{equation}
with
\begin{equation}
<\xi(t)> =0,
<\xi(t)\xi(t^{'})>=2\delta(t-t^{'}).
\end{equation}
Here, and throughout the text, $<...>$ correspond to ensemble averages.

For simplicity and convenience the equation is transformed into dimensionless
units\cite{Desloge} by setting $m=1$, $V_0=1$, $k=1$, with reduced variables
denoted again by the same symbols. The Langevin equation therefore takes the
form
\begin{equation}
\frac{d^{2}x}{dt^{2}}=-\gamma(x)\frac{dx}{dt}
-\frac{\partial V(x)}{\partial x} +F(t)+\sqrt{\gamma(x) T}\xi(t),
\end{equation}
where the potential is reduced to $V(x)=-\sin(x)$ and
\begin{equation}
\gamma(x)=\gamma_0(1-\lambda\sin(x+\theta)).
\end{equation}
Equation (2.4) is numerically solved (i.e., integrated using Ito definition)
to obtain the trajectories $x(t)$ of the particle for various initial 
conditions using Heun's method\cite{Nume,SRS,Mannella}. For initial positions 
$x(t=0)$ the period $-\frac{\pi}{2}<x\leq\frac{3\pi}{2}$ is divided uniformly 
into either $n=100$ or, in some cases, 200 values (and hence $n$ initial 
positions) and the initial velocity $v(t=0)$ is set equal to zero throughout in
the present work.

For each trajectory, corresponding to one initial position $x(0)$, the thermodynamic 
work done by the field $F(t)$ on the system (or the input energy) is
calculated using stochastic energetics formulation of Sekimoto as \cite{Sekimoto}:
\begin{equation}
W(0,N\tau)=\int_0^{N\tau}\frac{\partial U(x(t),t)}{\partial t}dt,
\end{equation}
where $N$ is a large integer denoting the number of periods taken to reach the
final point of the trajectory. The effective potential $U$ is given by
\begin{equation}
U(x(t),t)=V(x(t))-x(t)F(t).
\end{equation}
The mean input energy per period, for a given trajectory, is given as
\begin{equation}
\overline{W}=\frac{1}{N}W(0,N\tau).
\end{equation}

Since $V(x)$ is not explicitly dependent on time, all of the contribution to 
$W(0,N\tau)$ comes from the second term in $U$, Eq. (2.7). 
\begin{equation}
W(0,N\tau)=-\int_0^{N\tau}{x(t)\frac{\partial F(t)}{\partial t}}dt,
\end{equation}
Note that this relation is true even if a constant load is applied, replacing 
$V(x) = - \sin(x)$ by $V(x) = -\sin(x) + xL$. That is the input energy with load 
$W(0,N\tau,L)$ is equal to the input energy without load $W(0,N\tau,0)$: 
\begin{equation}
W(0,N\tau,0) = W(0,N\tau,L). 
\end{equation}
However, though
$W(0,N\tau)$ does not explicitly depend on $V(x)$, the latter implicitly 
affects $W(0,N\tau)$ through $x(t)$. Because of the nonzero phase difference 
between $F(t)$ and $x(t)$ the integral $W(0,N\tau)$ is finite. The
nonzero phase difference results in hysteresis $\overline{x(F)}$, whose area
is a measure of energy dissipation per period by the system. The overall mean
input energy per period, $<\overline{W}>$, is calculated as an (ensemble) 
average over all the $n$ trajectories (corresponding to all the initial 
positions considered). Again, we find that the mean hysteresis loop area 
$<\overline{A}>=<\overline{W}>$. Similarly, the mean velocity $\overline{v}$, 
over one trajectory, is calculated as
\begin{equation}
\overline{v}=\frac{1}{N\tau}(x(t=N\tau) - x(t=0)),
\end{equation}
and the overall mean (net) velocity or the ratchet current, $<\overline{v}>$,
is calculated as the ensemble average over all the $n$ trajectories. Typically,
the value of $N$ is chosen to have relatively small error bars so that the 
qualitative features of the results are not obscured.

\section{Nonuniform friction and ratchet effect: A summary}

\subsection{Why a periodically varying friction leads to ratchet current?}

In this subsection, we present a qualitative discussion on why a spatially 
varying friction coefficient of the medium whose periodicity is same as that of 
the underlying potential function may lead to net asymmetric current when
driven by a temporally symmetric periodic force field.

\begin{figure}[htp]
\centering
\includegraphics[width=16cm,height=20cm]{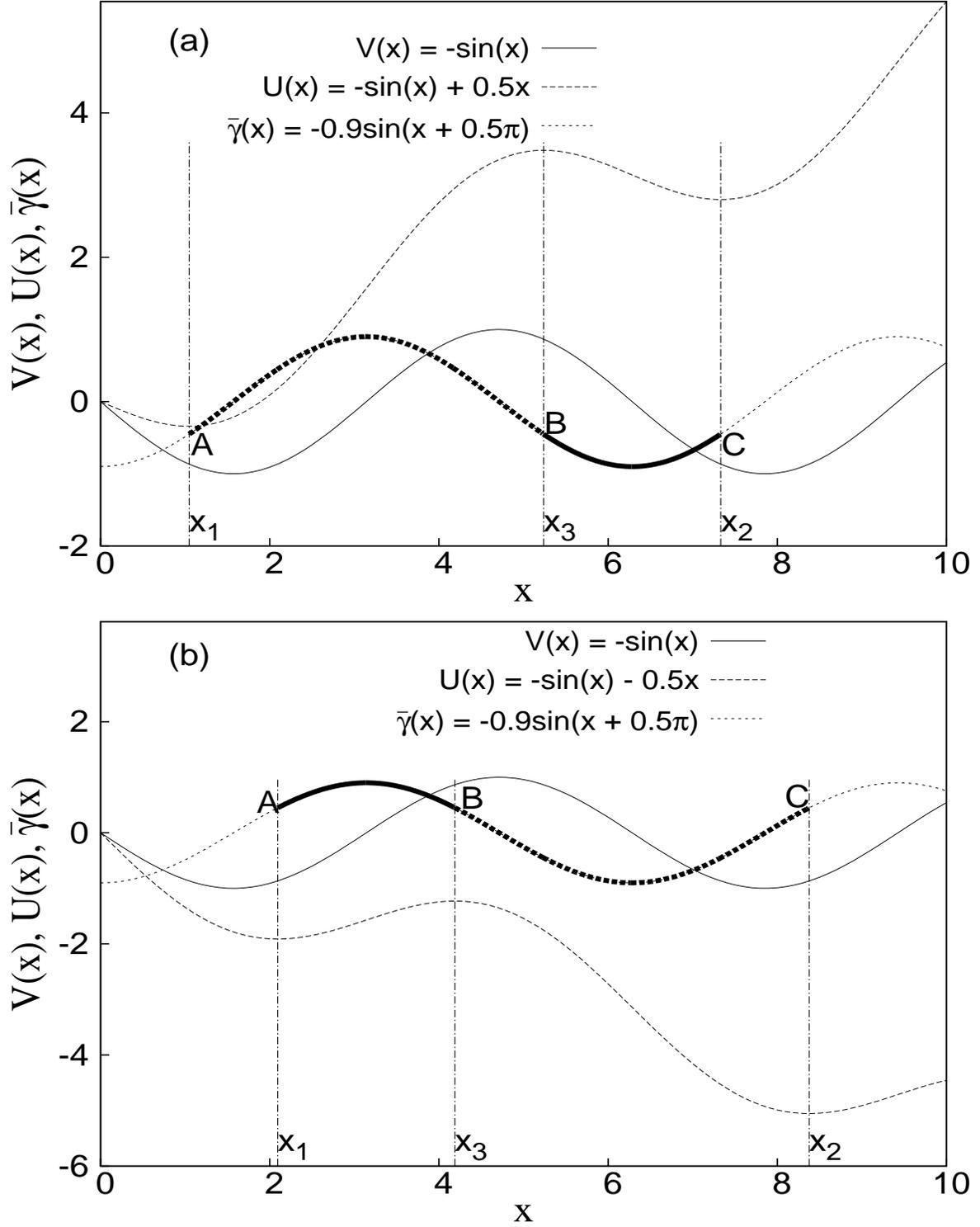}
\vspace*{-5mm}
\caption{The figure shows the variation of $V(x), U(x), \overline{\gamma}(x)$ 
as a function of $x$. Here, ${\theta} = 0.5{\pi}$, ${\lambda} = 0.9$ with 
$F_0 = -0.5$ in (a) and $0.5$ in (b).}
\end{figure}

\begin{figure}[htp]
\centering
\includegraphics[width=16cm,height=10cm]{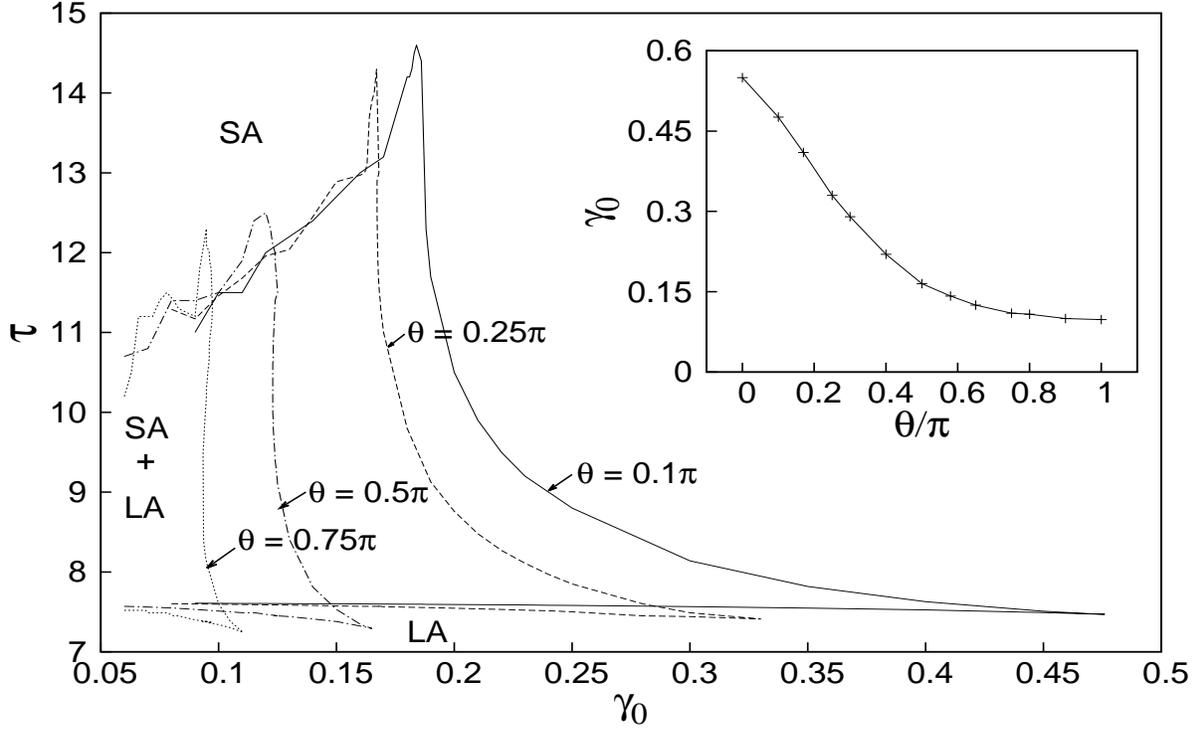}
\vspace*{-5mm}
\caption{The figure shows the boundaries of regions of coexistence of SA and 
LA states for $\theta = 0.10\pi, 0.25\pi, 0.5\pi$ and $0.75\pi$ at 
$T=0.000001$. In the inset is plotted, for various $\theta$, the value of 
$\gamma_0$ where the two dynamical states are not discernible.}
\end{figure}

\begin{figure}[htp]
\centering
\includegraphics[width=16cm,height=20cm]{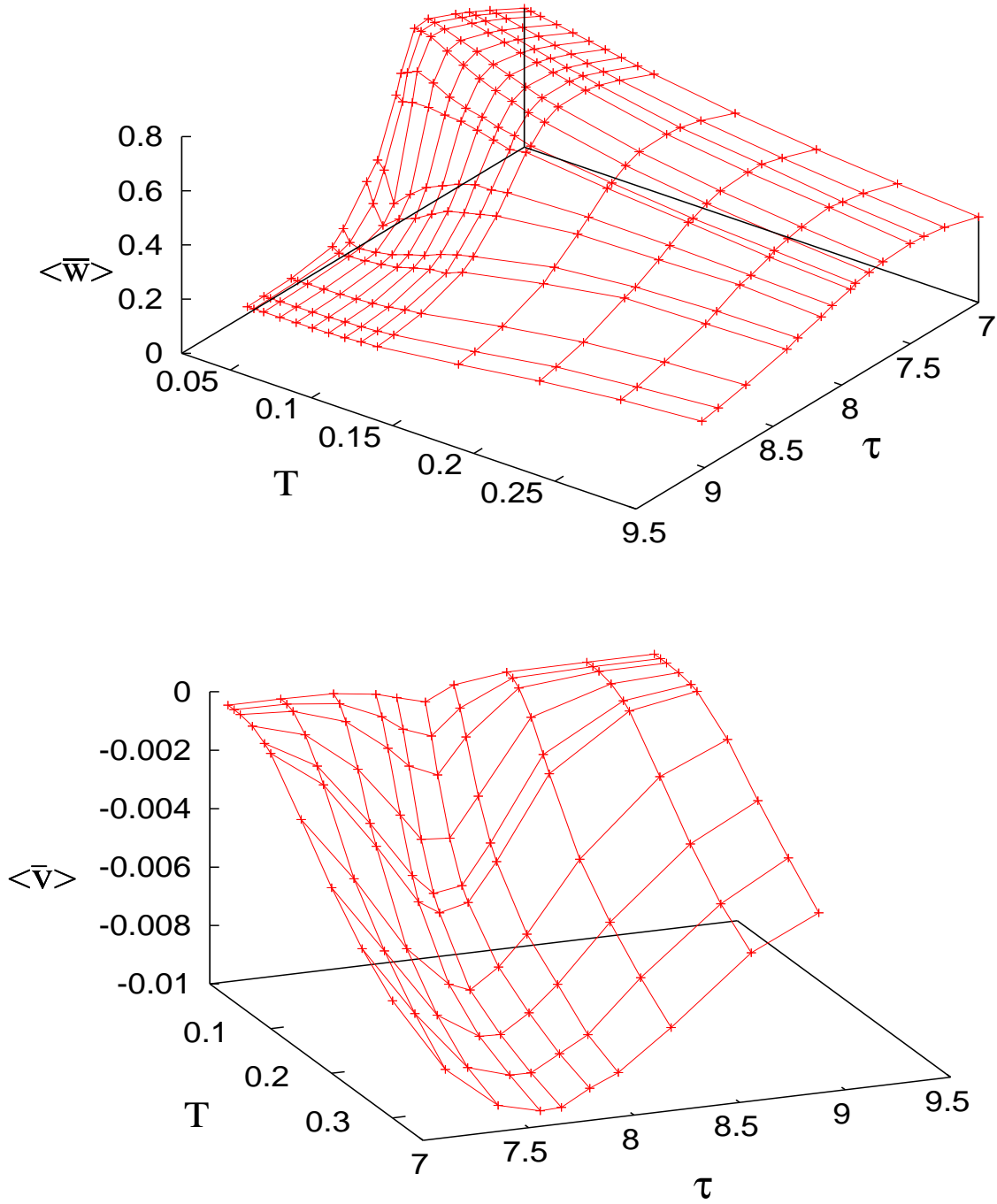}
\vspace*{-5mm}
\caption{The top figure shows the variation of $<\overline{W}>$ as a function 
of forcing period $\tau$ and temperature $T$. Here ${\theta} = 0.5{\pi}$, 
${\lambda} = 0.9$ , $F_0 = 0.2$ and $\gamma_0$ = $0.07$. For the same set of 
parameter values, the bottom figure shows the variation of $<\overline{v}>$ as 
a function of forcing period $\tau$ and temperature $T$.}
\end{figure}

\begin{figure}[htp]
\centering
\includegraphics[width=16cm,height=10cm]{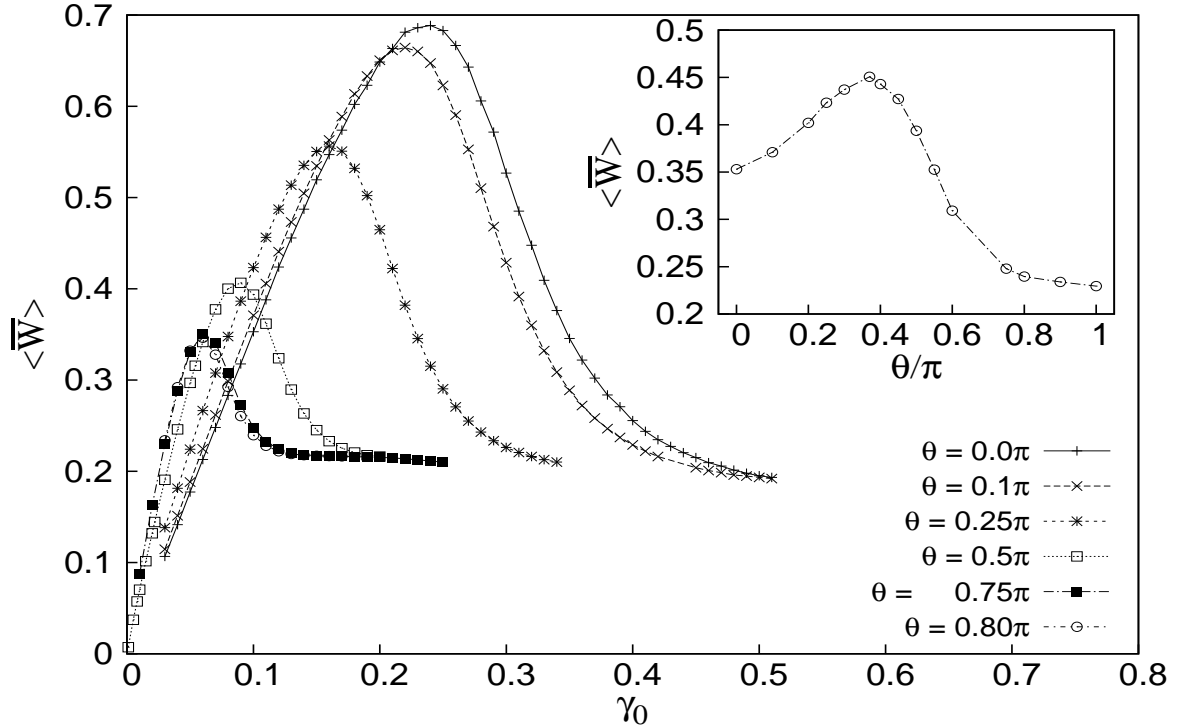}
\vspace*{-5mm}
\caption{The figure shows the variation of $<\overline{W}>$ as a function of 
$\gamma_0$. Here ${\lambda} = 0.9$ , $F_0 = 0.2$, $T = 0.1$ and $\tau$ = $8.0$ 
for a set of $\theta$ values as indicated in the plot. In the inset is shown the 
variation of $<\overline{W}>$ as a function of $\theta$ for an $intermediate$ value
of $\gamma_0 = 0.10$, keeping other parameters same as that of the main plot.}
\end{figure}

\begin{figure}[htp]
\centering
\includegraphics[width=16cm,height=10cm]{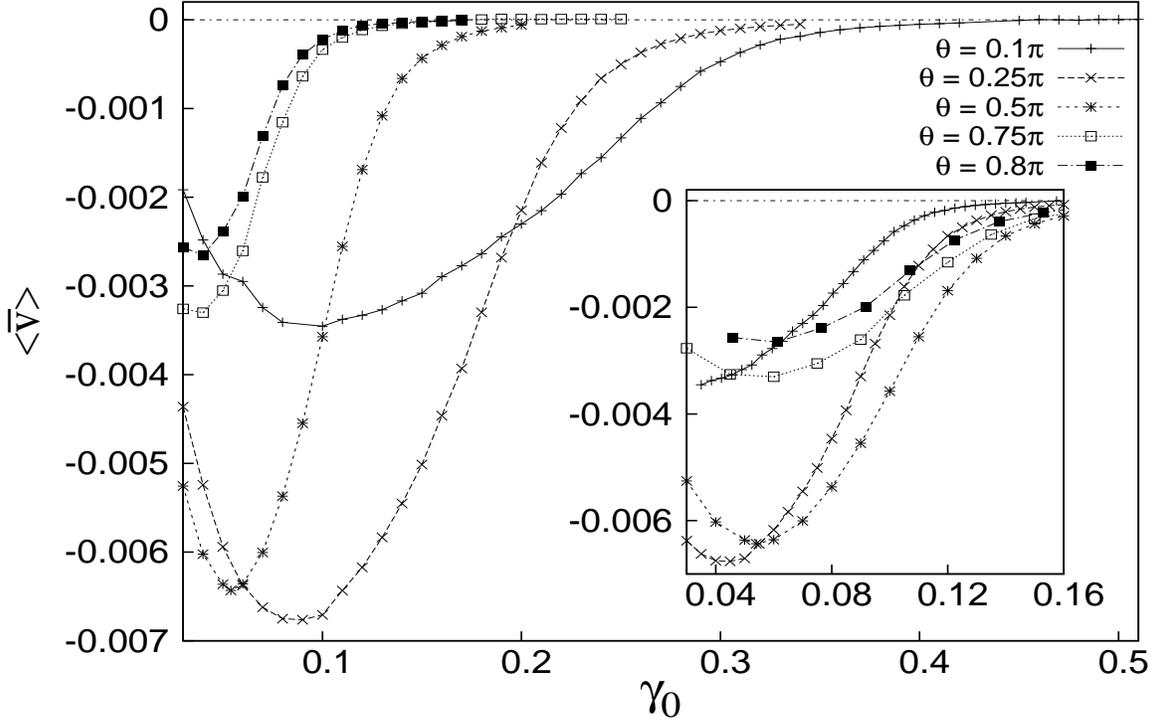}
\vspace*{-5mm}
\caption{The figure shows the variation of $<\overline{v}>$ as a function of 
$\gamma_0$. Here ${\lambda} = 0.9$ , $F_0 = 0.2$, $T = 0.1$ and $\tau$ = $8.0$ 
for a set of $\theta$ values as indicated in the plot. In the inset is shown 
the same plot  but after a suitable scaling factor (different for different
$\theta$, see text) is applied to the abscissa.}
\end{figure}

\begin{figure}[htp]
\centering
\includegraphics[width=16cm,height=10cm]{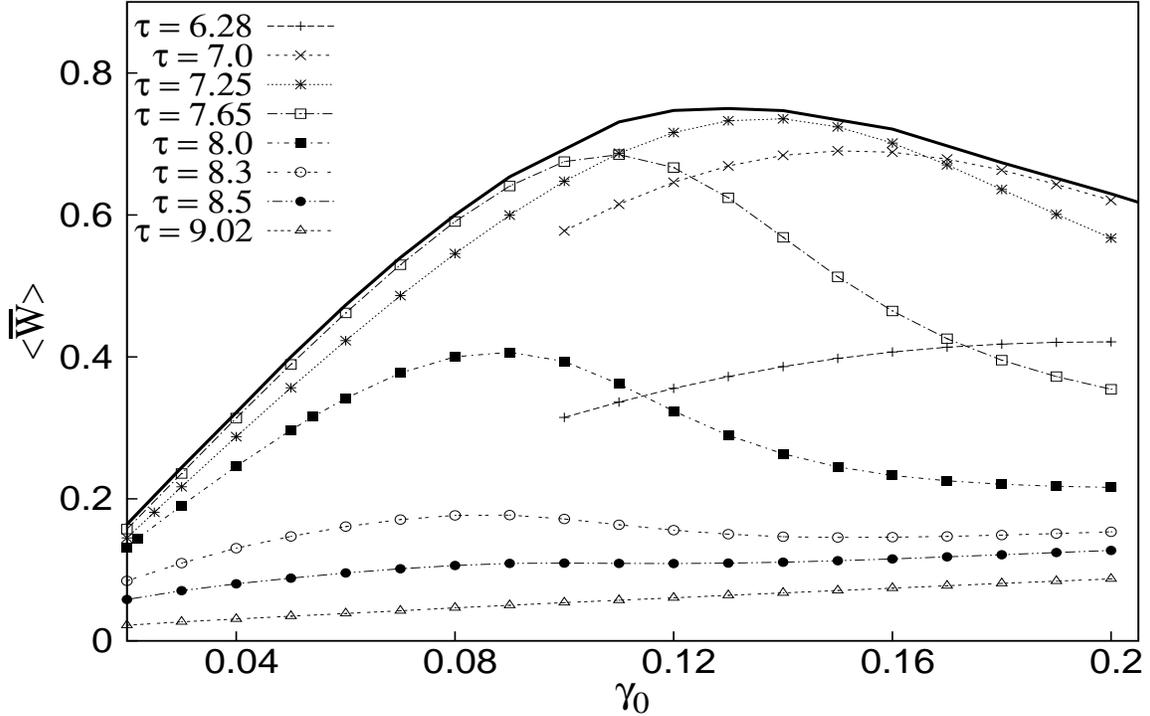}
\vspace*{-5mm}
\caption{The figure shows the variation of $<\overline{W}>$ as a function of 
$\gamma_0$. Here ${\lambda} = 0.9$ , $F_0 = 0.2$, $T = 0.1$ and 
$\theta$ = $0.5\pi$ for a set of $\tau$ values as indicated in the plot. The 
thick line however denotes the maximum $<\overline{W}>$, which occurs for 
different driving frequencies $\omega$, as a function of $\gamma_0$.}
\end{figure}

\begin{figure}[htp]
\centering
\includegraphics[width=16cm,height=10cm]{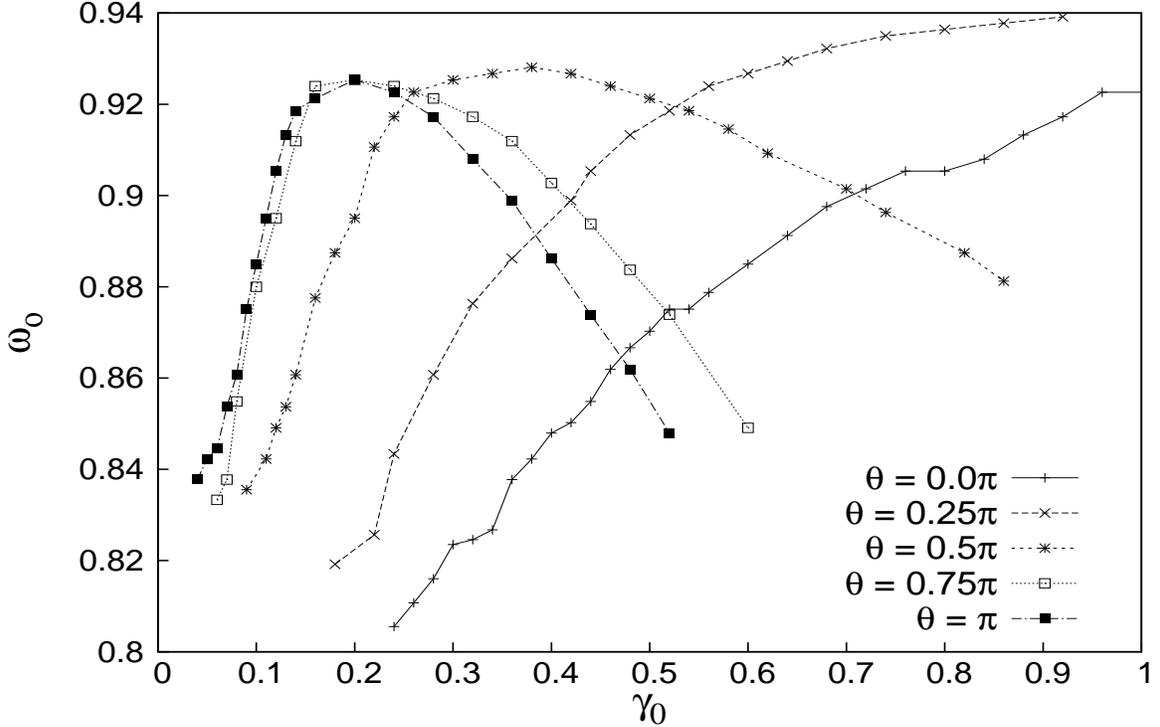}
\vspace*{-5mm}
\caption{The figure shows the conventional  resonance  frequency $\omega_0$ as 
a function of $\gamma_0$. Here ${\lambda} = 0.9$, $F_0 = 0.2$ and $T = 0.1$ 
for a set of $\theta$ values as indicated in the plot.}
\end{figure}

\begin{figure}[htp]
\centering
\includegraphics[width=16cm,height=10cm]{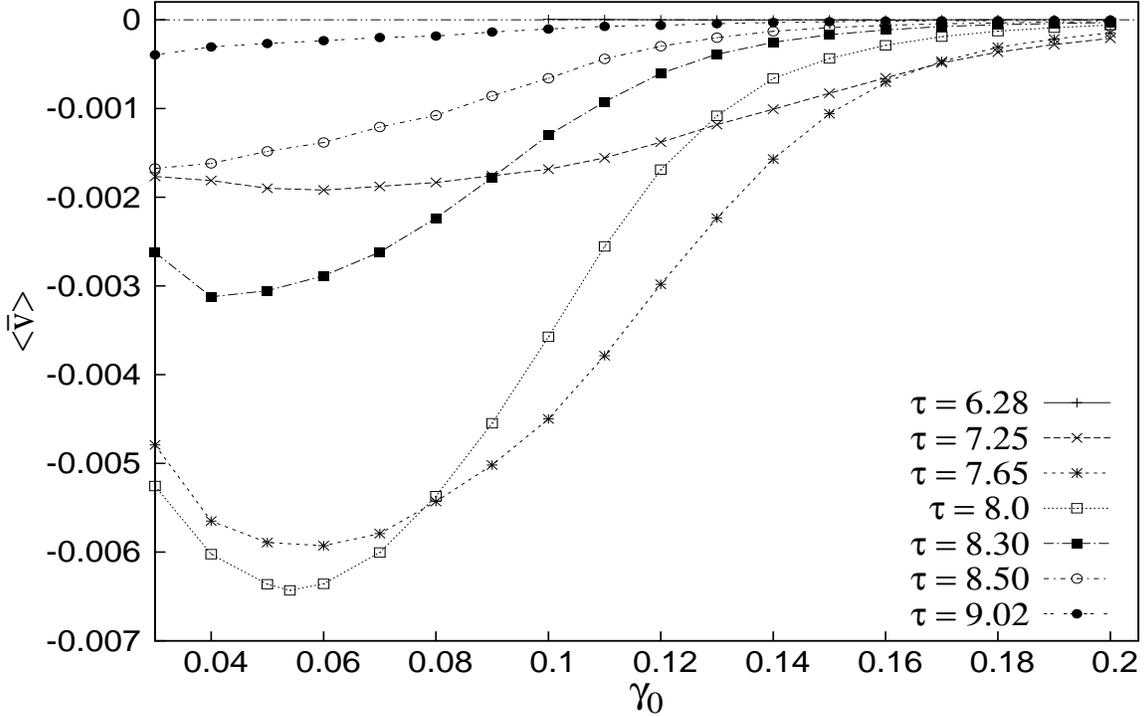}
\vspace*{-5mm}
\caption{The figure shows the variation of $<\overline{v}>$ as a function of 
$\gamma_0$. Here ${\lambda} = 0.9$ , $F_0 = 0.2$, $T = 0.1$ and 
$\theta$ = $0.5\pi$ for a set of $\tau$ values as indicated in the plot.}
\end{figure}

\begin{figure}[htp]
\centering
\includegraphics[width=16cm,height=10cm]{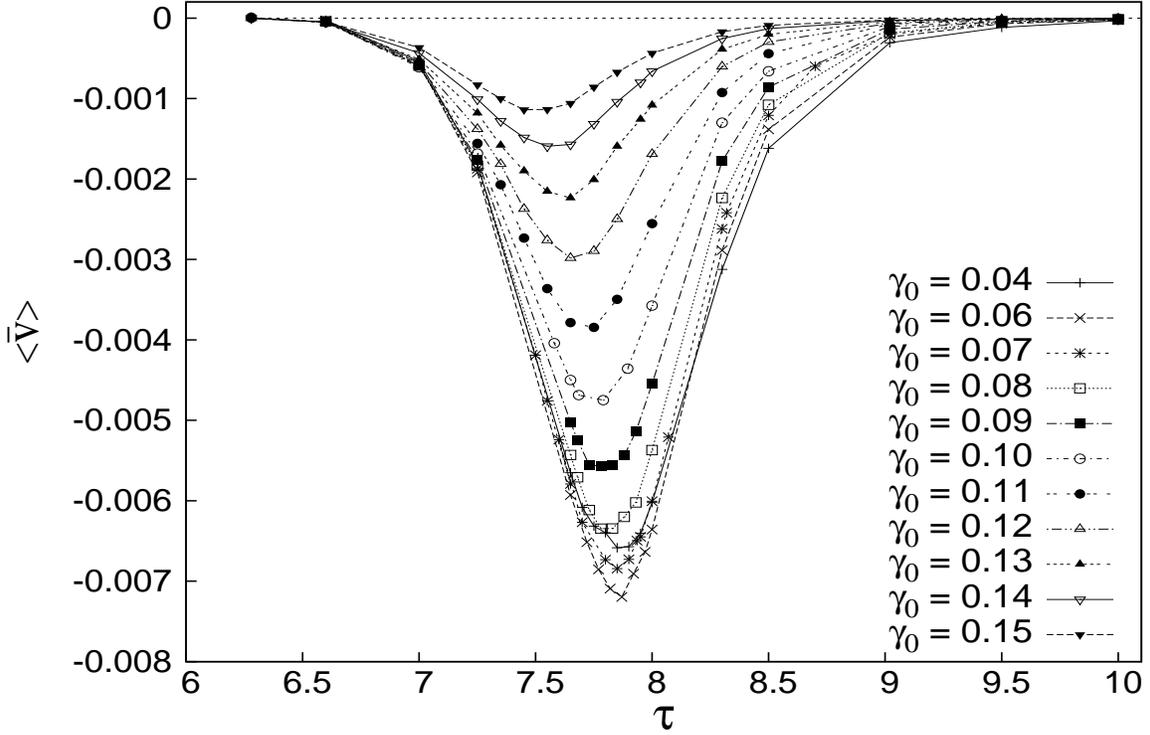}
\vspace*{-5mm}
\caption{The figure shows the variation of $<\overline{v}>$ as a function of 
$\tau$. Here ${\lambda} = 0.9$, $F_0 = 0.2$, $\theta$ = $0.5\pi$ and $T = 0.1$ 
for a set of $\gamma_0$ values as indicated in the plot.}
\end{figure}

\begin{figure}[htp]
\centering
\includegraphics[width=16cm,height=12cm]{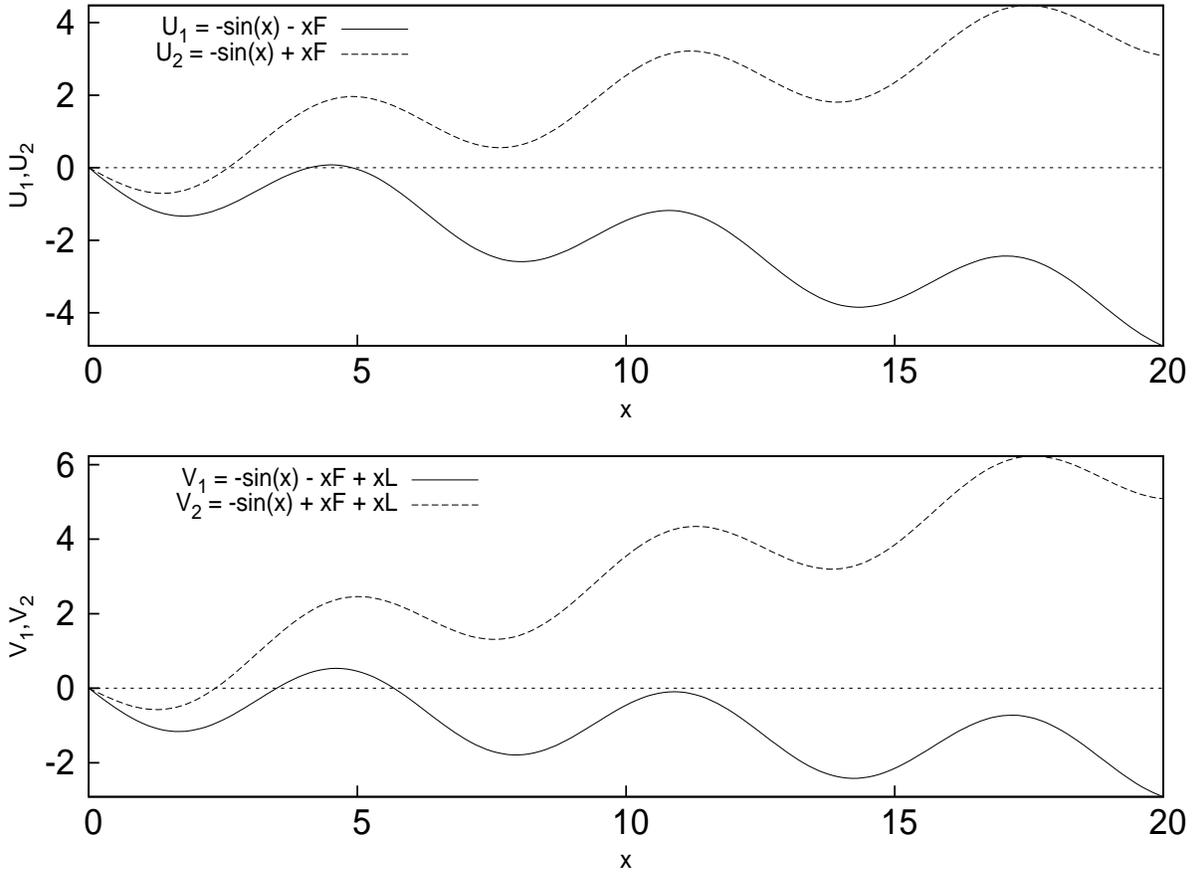}
\vspace*{-5mm}
\caption{The panels shows the effect of the forcing on the potential.
At the top panel is shown (in the absence of load)
$U_1$ as the effective potential at the beginning of the forcing period whereas $U_2$
is the effective potential at half-period of the forcing.
The bottom panel shows the effect of the forcing on the potential when a load is present.
This is the so called washboard potential. $V_1$ is the effective potential
at the beginning of the forcing period whereas $V_2$ is the effective potential at
 half-period of the forcing.}
\end{figure}

\begin{figure}[htp]
\centering
\includegraphics[width=16cm,height=10cm]{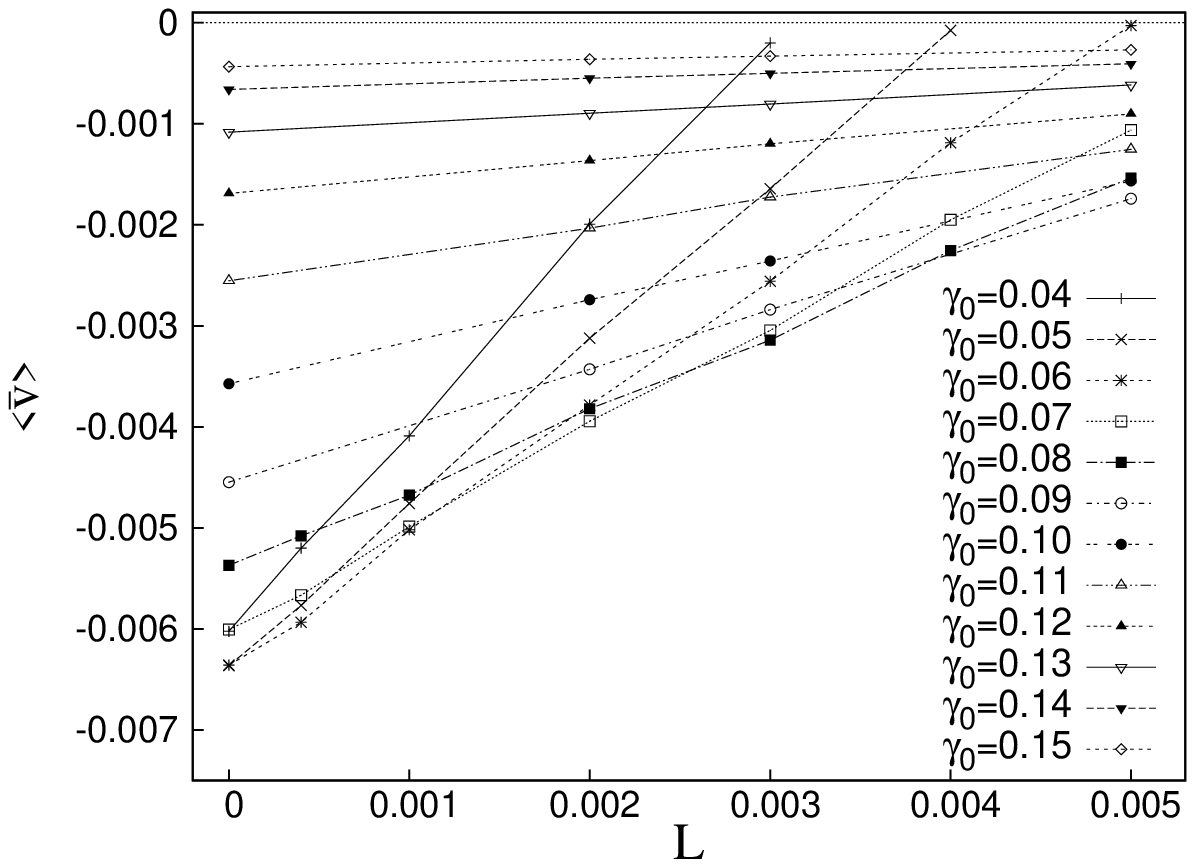}
\vspace*{-5mm}
\caption{The figure shows the variation of $<\overline{v}>$ as a function of 
load $L$ for a set of $\gamma_0$ values as indicated in the graph. Here 
${\lambda} = 0.9$, $F_0 = 0.2$, $\theta$ = $0.5\pi$, $T = 0.1$ and 
$\tau = 8.0$. The top most horizontal line is the zero line.}
\end{figure}

\begin{figure}[htp]
\centering
\includegraphics[width=16cm,height=10cm]{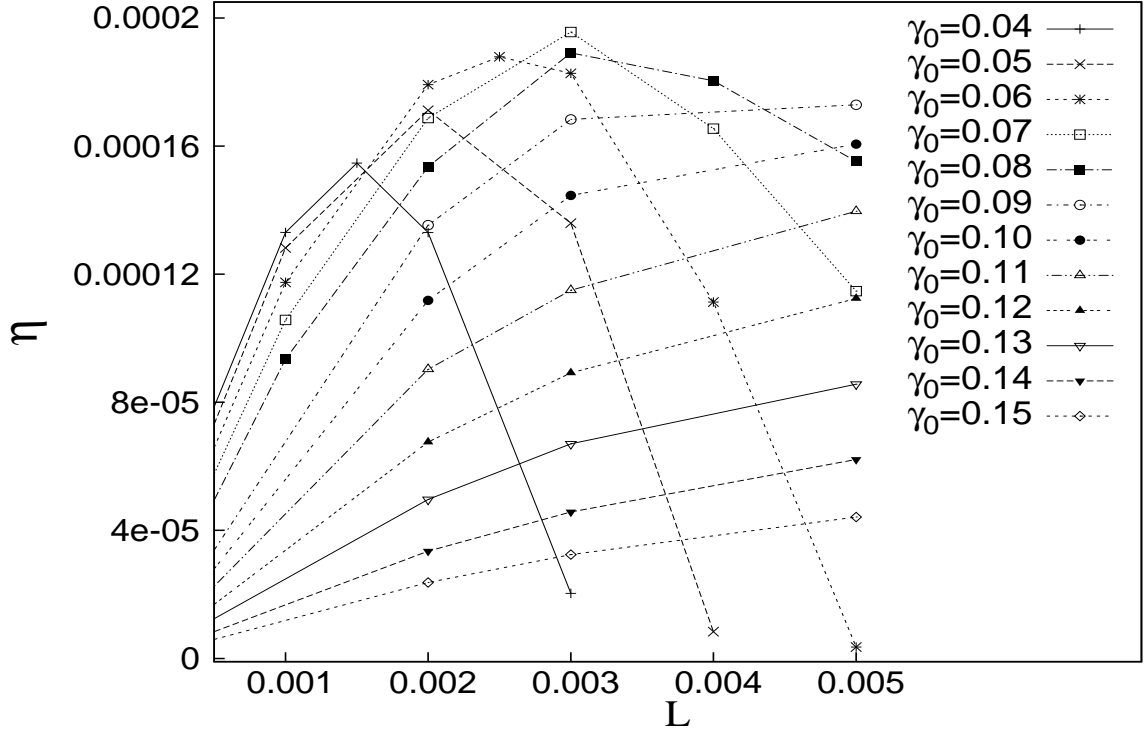}
\vspace*{-5mm}
\caption{The figure shows the variation of $\eta$ as a function of load $L$ 
for a set of $\gamma_0$ values as indicated in the graph. Here 
${\lambda} = 0.9$, $F_0 = 0.2$, $\theta$ = $0.5\pi$, $T = 0.1$ and 
$\tau = 8.0$.}
\end{figure}

Figure 1 shows three curves: (1) the sinusoidal potential $V(x)=-\sin x$, (2)
the effective potential $U(x)=V(x)-xF(t)$, where the external drive 
$F(t)=F_0\cos \omega t$ when (a) $F(t)=-F_0=-0.5$ and (b) $F(t)=F_0=0.5$, and
(3) $\overline{\gamma}(x)=\gamma(x)-1$, where the friction coefficient
$\gamma(x)=\gamma_0(1-0.9\sin (x+\theta))$, with $\theta=0.5\pi$. We plot 
$\overline{\gamma}(x)$ instead of $\gamma(x)$ (here with $\gamma_0=1.0$) for 
convenience of comparison. Notice that $U(x)$ with $F(t)$ = (a) -0.5 and (b) 
+0.5, provide exactly the same potential barrier but in reverse directions. 
Therefore, if $\gamma(x)$ were uniform or nonuniform with $\theta=0$ or $\pi$ 
there would be no net particle current because particle would be equally 
likely to move either to the right or to the left. However, if 
$\theta\neq 0,~\pi$ the left-right symmetry is broken, that is, the particle 
encounters different situations in the two directions. 

We focus attention on a period of $\overline{\gamma}(x)$ with end points 
$x=x_1$ and $x=x_2$ corresponding to the two consecutive minima of the 
effective potential $U(x)$ with intervening maximum at $x=x_3$. AB denotes the 
segment of $\overline{\gamma}(x)$ from $x_1$ to $x_3$ and BC from $x_3$ to 
$x_2$. In Fig. 1a with $F(t)=-F_0=-0.5$ in the segment BC the mean value of 
$\overline{\gamma}(x)$ (or that of $\gamma(x)$) is smaller than in the segment 
AB. Also, since in this case mean slope of $U(x)$ is positive it is more 
likely for the particle to move to the left (from $x_3$ towards $x_1$). In 
Fig. 1b the situation is exactly the reverse. However, from Fig. 1a it 
is clear that in the former case the particle encounters a smaller barrier to 
surmount in the left direction with smaller mean friction (segment BC) and a 
larger barrier in the right direction with larger mean friction (segment AB). 
On the other hand in the latter case of Fig 1b the particle is to surmount the 
smaller barrier (of same height as in Fig. 1a) in the right direction but with 
a larger mean friction (segment AB) and the larger barrier in the left 
direction with a smaller mean friction (segment BC). Since the temperature is 
not large (typically $<0.2$) it is the smaller barrier that should contribute 
more to the dispacement than the larger barrier and of course a smaller 
friction (and hence larger mobility) is more helpful in barrier crossing. 
Therefore one would expect a net current in the left direction when 
$F(t)=F_0\cos \omega t$ is considered over the entire period 
$\tau=2\pi/\omega$. And this explains the net negative current observed in our 
numerical results to be described later for all $0<\theta<\pi$. However, it is 
to be noted that the values of the parameters $F_0$ and $\gamma_0$ used here 
for illustration are to exaggarate the situation to bring the point home. In 
our work, in the following, we keep $F_0=0.2$ and $\gamma_0$ is varied but rarely 
so large as 1.0 used here. Yet the argument remains qualitatively valid. Can 
one extend this analogous argument to explain the magnitude of the net current 
too? 

Of course, this situation can be brought out more effectively when the 
frictional asymmetry with respect to $V(x)$ is maximum at $\theta=0.5\pi$ and
thus one would expect the ratchet current to be the largest for 
$\theta=0.5\pi$. However, frictional asymmetry is not the only factor that
comes into play. A closer look at the nature of particle trajectories in the
($\gamma_0-\tau$) space, as has been obtained previously is a similar system as
in the present case, is quite educative in this respect.

\subsection{A brief summary of results from earlier works}

At low temperatures the system shows two and only two kinds of particle 
trajectories, $x(t)$: a large amplitude trajectory (LA) with large phase lag 
with respect to $F(t)$ and another with small amplitude (SA) and also small 
phase lag\cite{Saikia3,Wanda2,Donrich}. The two states of trajectories occur
in a restricted range of parameter space $(\gamma,\omega)$. In particular $\omega$ 
should be close to the natural frequency of oscillation $(\omega\approx 1)$ at 
the bottom of the potential. This frequency $(\omega\approx 1)$ is much larger 
(about 10$^2$ times) than the mean rate of passage across the potential barrier.
The occurrence of these trajectories depends on the 
initial conditions of position $x(t=0)$ (within a period of $V(x)$) and 
velocity $v(t=0)$ (Here we choose $v(0)=0$.). Figure 2 is an extension of 
Fig. 7 of Ref.\cite{Donrich} showing the region of coexistence of these two 
`dynamical states' of trajectories in the ($\gamma_0-\tau$) plane. The nature 
of the curves for various $\theta$ values clearly suggests that the 
effectiveness of $\gamma_0$ decreases as $\theta$ is decreased\cite{Donrich}. 
This can also be seen from Fig 1a that when, for example, $\theta=0~(\pi)$ the 
particle encounters the least (largest) average friction while having its 
motion about the bottom of the potential. The value of $\gamma_0$ at which the 
region of coexistence of the two states of trajectories vanishes may roughly 
be taken as the reciprocal measure of an effective $\gamma_0$ for a given 
$\theta$. In the inset of Fig.2 are plotted the $\gamma_0$ values at which the 
regions of coexistence just vanish. When the friction becomes less effective 
(smaller $\theta$) the net particle current may be larger than when the 
friction becomes more effective (larger of $\theta$ in the range 
$0<\theta<\pi$). Thus the frictional asymmetry plays a dual role as far as 
the magnitude of the net (ratchet) current is concerned. 

We confine our work mostly in the region of coexistence of the two dynamical
states for we expect to get interesting results in this region. In particular,
it has been shown that the energy of dissipation of the system per period 
$\tau$ of $F(t)$ shows a maximum at an intermediate noise strength (or 
temperature $T$). Or, equivalently, the ratio of the output signal to input 
noise peaks as a function of temperature for a given $\tau,~\gamma_0$, and small
$F_0$. This important phenomenon is termed as stochastic resonance. Stochastic
resonance is usually observed in bistable systems at a frequency of the order
 of the mean rate of passage across the barrier. However, in this case it
refers to the sinusoidal potential $U(x)$. The energy dissipation or the input 
energy plays a key role in the efficiency calculation of the frictional 
ratchet as will be discussed in the subsequent sections. The input energy (and 
ratchet currents) were calculated earlier in the ($\tau-T$) 
plane\cite{Donrich}. The results are summarized in Fig.3. 

\section{Numerical results}

In the present numerical work, we take N = 200000 cycles of the drive to obtain
the necessary ratchet currents. We calculate the difference
$(x(t=N\tau) - x(t=0))$ for each initial condition used to calculate 
$\overline{v}=\frac{(x(t=N\tau)- x(t=0))}{N\tau}$. The deviations of these $\overline{v}$
from the mean ${<\overline{v}>}$ is calculated as an ensemble average over all 
trajectories whose initial positions are uniformly distributed within a period of the 
potential. One can then obtain the necessary standard 
deviations $\Delta v$ to form the error bars.

\subsection{Input energy and ratchet current}

We explore the behavior of net (ratchet) current, $<\overline{v}>$, as well as 
the input energy, $<\overline{W}>$, as a function of $\gamma_0$ for various
values of $\theta$ and $\tau$. From symmetry arguments $<\overline{v}>=0$ for
$\theta=0,~n\pi$ for all integral $n$. Also, as it should, $\pi\leq\theta<2\pi$
gives the same information about $<\overline{v}>$ as in case of 
$0\leq\theta<\pi$, for any value of $\gamma_0,~\lambda$ and $\tau$, only the 
direction is reversed. Note that we keep the amplitude $F_0$ of the drive 
$F(t)$ fixed and equal to 0.2. Though $<\overline{v}>=0$ for $\theta=0,~\pi$, 
the input energy $<\overline{W}>\neq 0$. Figure 4 shows $<\overline{W}>$ as a 
function of $\gamma_0$ at a constant temperature $T=0.1$ and $\tau=8.0$ for 
various values of $\theta$. Interestingly, for small $\gamma_0<\gamma_{01}$ 
($\gamma_{01}\approx 0.05$) and for large $\gamma_0>\gamma_{02}$
($\gamma_{02}\approx 0.2$) , $<\overline{W}>(\theta)$ shows monotonic 
(respectively, increasing and decreasing) behavior for fixed $\gamma_0$. 
However, in the intermediate range of $\gamma_{01}\leq\gamma_0\leq\gamma_{02}$,
$<\overline{W}>(\theta)$ shows peaking behavior. The width of the 
intermediate range of $\gamma_0$ depends on $\tau$, etc. It is roughly in this 
intermediate range where the phenomenon of stochastic resonance (peaking of
$<\overline{W}>$ as a function of temperature for fixed $\gamma_0, \theta, 
F_0$ and $\tau$) is obtained. In the inset of Fig. 4, $<\overline{W}>(\theta)$
 is shown for one of such intermediate value of $\gamma_0 = 0.10$ 
which displays the described peaking behaviour.

In Fig. 5 is shown the ratchet current $<\overline{v}>$ as a function of 
$\gamma_0$ for various values of $\theta$ with other parameters fixed as in 
Fig. 4. Note that particle trajectory properties (including the hysteresis
loops) for $\theta=0.5\pi$ are closest to the properties of that of a system
with uniform friction with $\lambda=0$ (see Fig. 5 of Ref.\cite{Donrich}).
As mentioned earlier, $\theta=0.5\pi$ provides the largest left-right 
asymmetry due to frictional nonuniformity and hence one would expect to obtain
largest $<\overline{v}>$ for $\theta=0.5\pi$. However, Fig. 5 shows that for
$\gamma_0\geq 0.05$, the mean velocity $<\overline{v}>$ are larger for 
$\theta <0.5\pi$ than $<\overline{v}>$ for $\theta =0.5\pi$. 

The above apparent anomaly can be explained by the suggestion that 
$<\overline{v}>$ is due to the combined effect of frictional asymmetry and 
effective value $\overline{\gamma_0}$ of $\gamma_0$. The inset of Fig. 2 gives 
an idea that as $\theta$ decreases from $0.5\pi$, $\overline{\gamma_0}$ 
decreases and it gets enhanced $\overline{\gamma_0}$ for $\theta>0.5\pi$. For 
example, in Fig. 2, the region of coexistence of the dynamical states of 
trajectories closes at $\gamma_0=0.165$ for $\theta=0.5\pi$ and for 
$\theta=0.25\pi$ and $0.75\pi$ the regions close respectively at 
$\gamma_0=0.33$ and 0.11. If we rescale the abscissa of Fig. 5 accordingly, 
for example by a factor of $\frac{0.165}{0.33}$ for $\theta=0.25\pi$, what 
results is shown in the inset of Fig. 5. It too, however, does not entirely 
isolate the two effects, as can be inferred from the inset; one should change 
the value of $<\overline{v}>$ as well, for example by a factor of, say, 0.7. 
But, of course, we have no basis to arrive at the said factor. The message of
Fig. 5, therefore, remains that $<\overline{v}>$, unlike 
$<\overline{W}(\theta)>$, shows nonmonotonic behavior as a function of $\theta$
at fixed $\gamma_0$ all through the abssissa of Fig. 5.

\begin{figure}[htp]
\centering
\includegraphics[width=16cm,height=10cm]{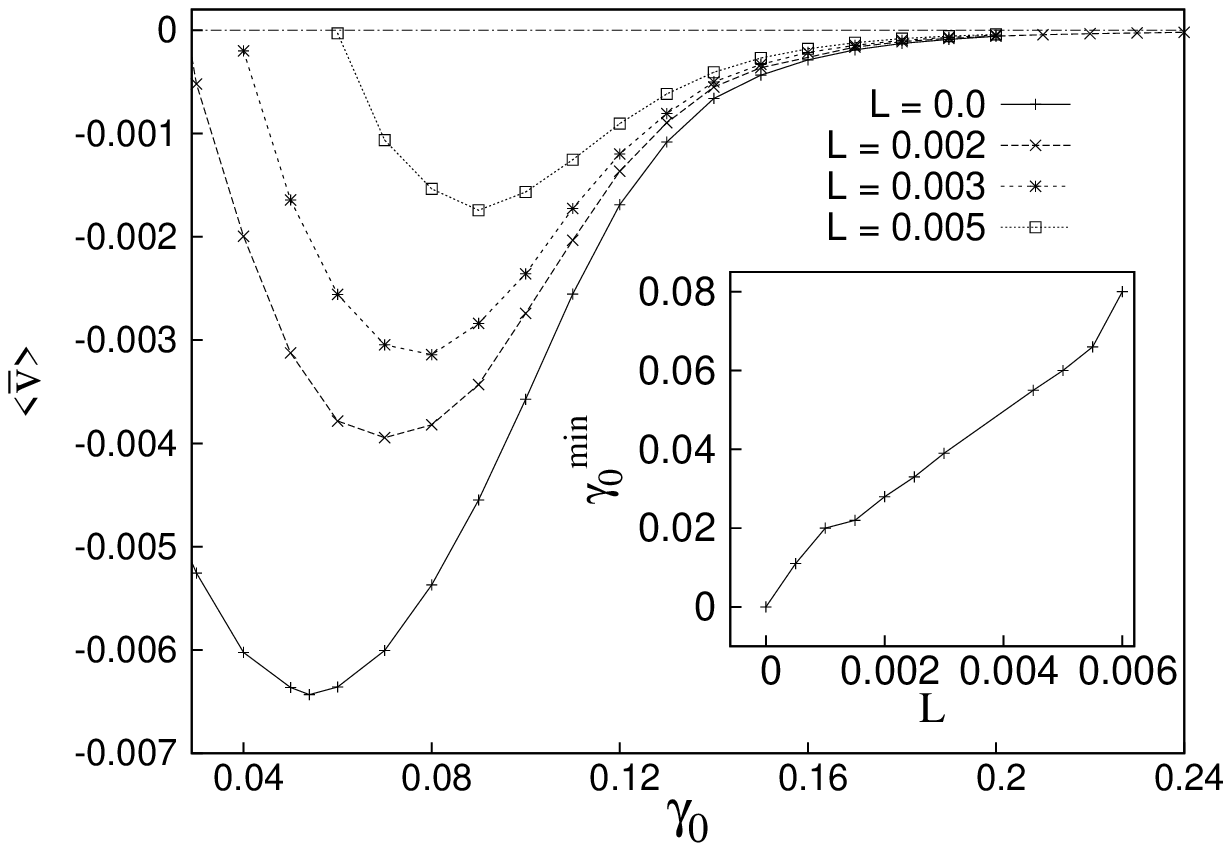}
\vspace*{-5mm}
\caption{The figure shows the variation of $<\overline{v}>$ as a function of 
$\gamma_0$ for the load-free case and also for a set of load, $L$, values as 
indicated in the plot. Here ${\lambda} = 0.9$, $F_0 = 0.2$, $\theta$ = 
$0.5\pi$, $T = 0.1$ and $\tau = 8.$ In the inset is calculated the maximum 
$\gamma_0 = \gamma_0^{min}$ where work can still be obtained by the ratchet as 
a function of load L.}
\end{figure}

Figure 6 gives the plot of $<\overline{W}>$ as a function of $\gamma_0$ for
$\theta=0.5\pi$ for various values of the period $\tau$ of the applied field
$F(t)$ at $T=0.1$. For other $\theta$ values the graph will have similar 
qualitative character. We have chosen $\theta=0.5\pi$ because it provides the 
largest asymmetry in addition to $\overline{\gamma_0}$ being equal to 
$\gamma_0$. From the graph it is clear that for all values of $\gamma_0$, the
input energy $<\overline{W}>$ shows peaking behaviour as a function of the
period $\tau$. However, the value of $\tau$, say, $\tau_0$, at which 
$<\overline{W}>$ peaks for a given value of $\gamma_0$ changes. The dark curve 
in Fig. 6 is the envelope of  $<\overline{W}>$ giving the largest possible
$<\overline{W}>(\gamma_0)$. For example at $\gamma_0 = 0.11$,  
$<\overline{W}>$ becomes largest for $\tau = \tau_0 = 7.4$, etc. Since 
$<\overline{W}>$ is equal to the hysteresis loop area it gives the mean energy 
absorbed per cycle of $F(t)$. Thus, $\tau_0$ can be considered as the period 
of conventional resonance at the given mean damping $\gamma_0$. The 
corresponding resonance frequencies $\omega_0=\frac{2\pi}{\tau_0}$ are plotted 
as a function of $\gamma_0$ in Fig. 7 for various values of $\theta$. The
curves show how damping affects the conventional resonance in a sinusoidal
potential, as opposed to in the usual limit of a parabolic potential. In 
particular, the curve for $\theta=0.5\pi$ provides important information about 
variation of resonace frequency, for example, of a driven damped (large 
amplitude) pendulum as a function of damping.

From the figure (Fig. 7) it is clear that the conventional resonance frequency
in a sinusoidal potential peaks at an intermediate value of 
$\gamma_0=\gamma_0^{peak}$. The value of $\gamma_0^{peak}$ increases with 
decreasing value of $\theta$. At this temperature ($T=0.1$) the numerical
results become less precise for smaller $\theta$ values. Also, one would expect
all curves to converge at a common value of resonance frequency $\omega_0$ for
all values of $\theta$ at $\gamma_0=0$. However, in our numerical work it is
hard to obtain the conventional resonance at low $\gamma_0$ values with small
error bars. That is why we have not shown the complete graphs in the lower
region of $\gamma_0$. Also, for smaller $\theta$ values (e.g., $\theta=0$ and
$0.25\pi$) the larger $\gamma_0$ region beyond $\gamma_0=1$ is not explored.    

In Fig. 8 is plotted $<\overline{v}>$ as a function of $\gamma_0$ for 
$\theta=0.5\pi$ and various values of $\tau$ as in Fig. 6. $<\overline{v}>$ 
vanishes for $\gamma_0=0$ for all $\tau$ and becomes very small for large
$\gamma_0$ values and hence peaks at intermediate $\gamma_0$ values. From the
graphs it is also clear that $<\overline{v}>$ shows nonmonotonic behaviour as a 
function of $\tau$ for fixed values of $\gamma_0$ in the entire range of 
abscissa. Figure 9 summarizes $<\overline{v}>(\tau)$ for various $\gamma_0$ 
values. Interestingly, the graphs show that appreciable ratchet current 
$<\overline{v}>$ can be obtained only in a limited range of $\tau$ values,
roughly $6.5\leq \tau\leq 9.5$, for all values of $\gamma_0$. Note that this is roughly
the domain of parameter space ($\gamma_0-\tau$) where stochastic
resonance is expected to occur in sinusoidal potentials. Just as the period of 
conventional resonance, the value of $\tau$ at which $<\overline{v}>$ peaks 
decreases with increase of $\gamma_0$. Although interesting, we cannot ascribe 
any physical significance to this behaviour unlike in the previous case of 
conventional resonance.

\begin{figure}[htp]
\centering
\includegraphics[width=16cm,height=10cm]{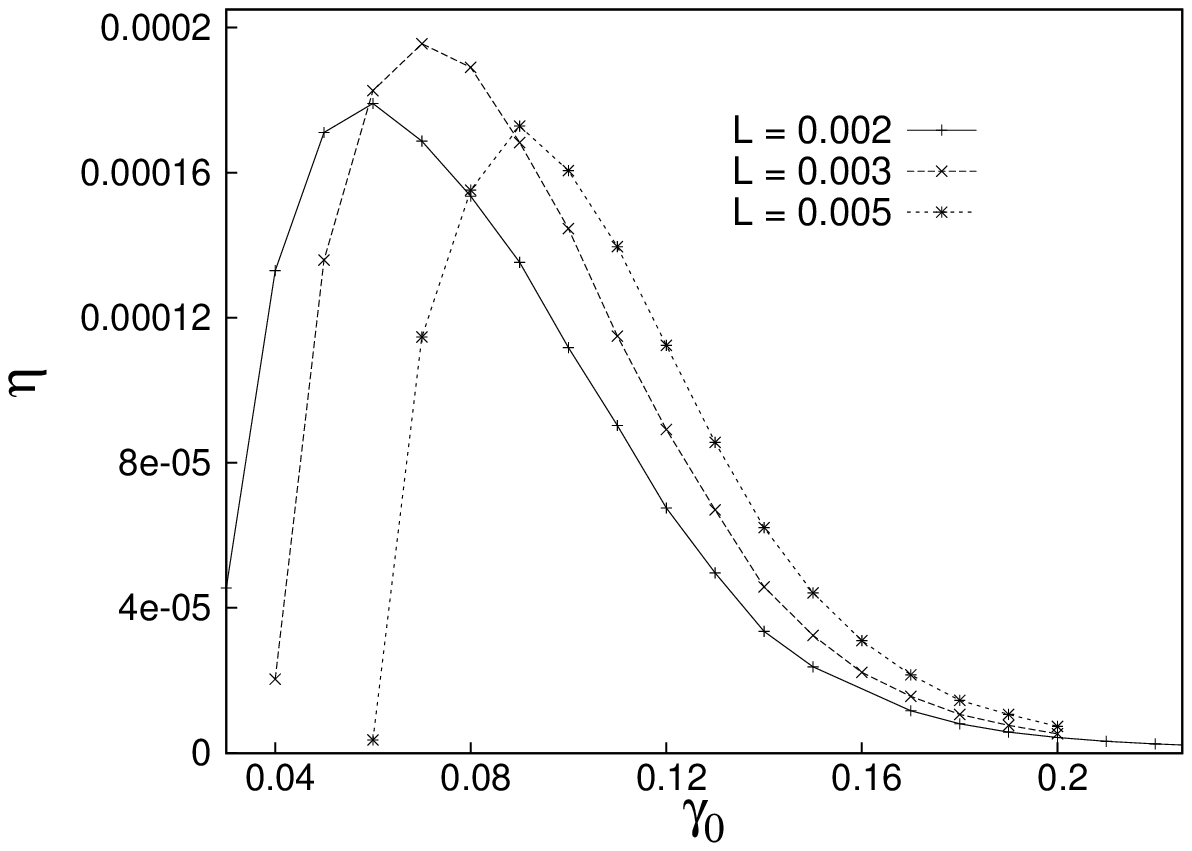}
\vspace*{-5mm}
\caption{The figure shows the variation of $\eta$ as a function of $\gamma_0$ 
for the set of load, $L$, values as indicated in the plot. Here ${\lambda} = 
0.9$, $F_0 = 0.2$, $\theta$ = $0.5\pi$, $T = 0.1$ and $\tau = 8.$}
\end{figure}

\begin{figure}[htp]
\centering
\includegraphics[width=16cm,height=10cm]{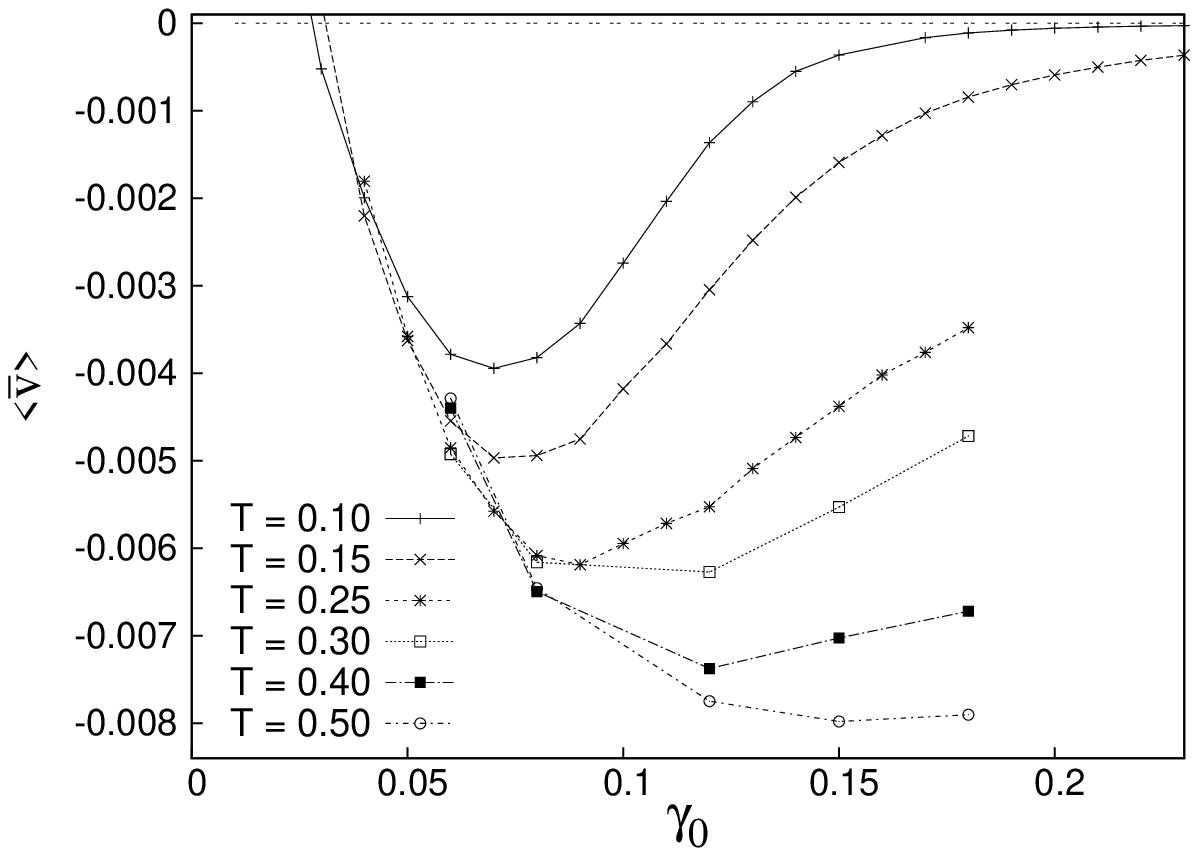}
\vspace*{-5mm}
\caption{The figure shows the variation of $<\overline{v}>$ as a function of 
$\gamma_0$ for various $T$ values as indicted in the graph for constant load 
$L = 0.002$ and $\tau = 8$. Here ${\lambda} = 0.9$, $F_0 = 0.2$ and $\theta$ = 
$0.5\pi$}
\end{figure}

\begin{figure}[htp]
\centering
\includegraphics[width=16cm,height=10cm]{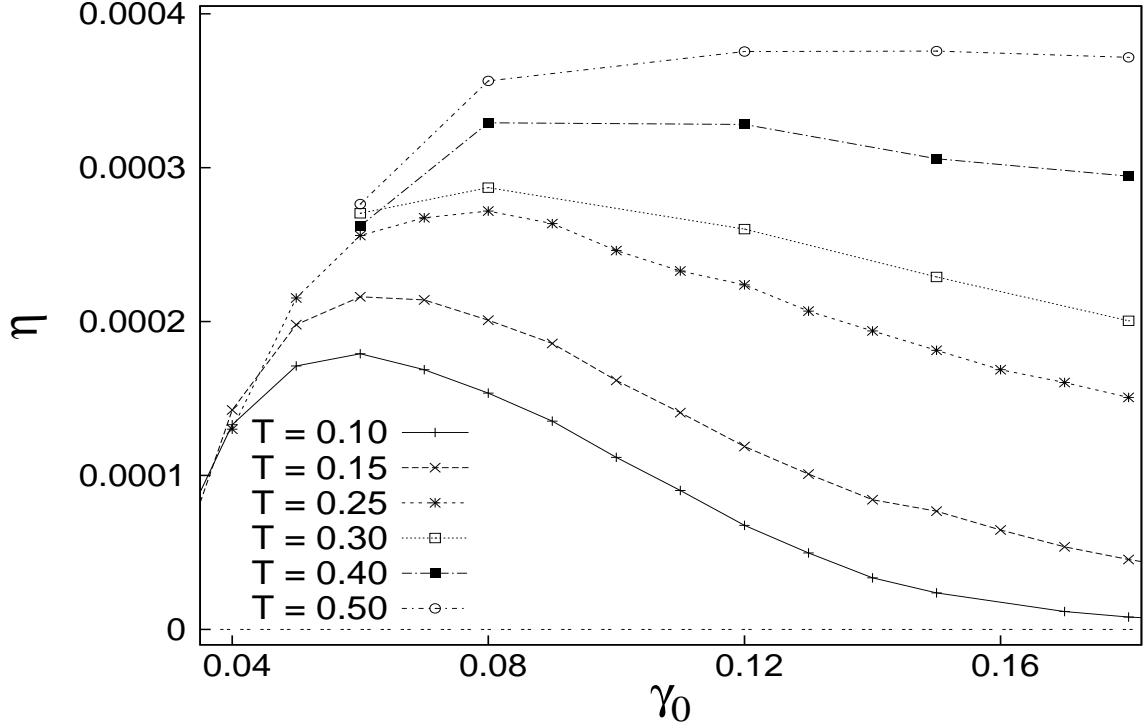}
\vspace*{-5mm}
\caption{The figure shows the variation of $\eta$ as a function of $\gamma_0$ 
for various $T$ values (as indicated in the bottom left region of the graph) 
for constant load $L = 0.002$ and $\tau = 8$. Here ${\lambda} = 0.9$, $F_0 = 
0.2$ and $\theta$ = $0.5\pi$.}
\end{figure}

\begin{figure}[htp]
\centering
\includegraphics[width=16cm,height=10cm]{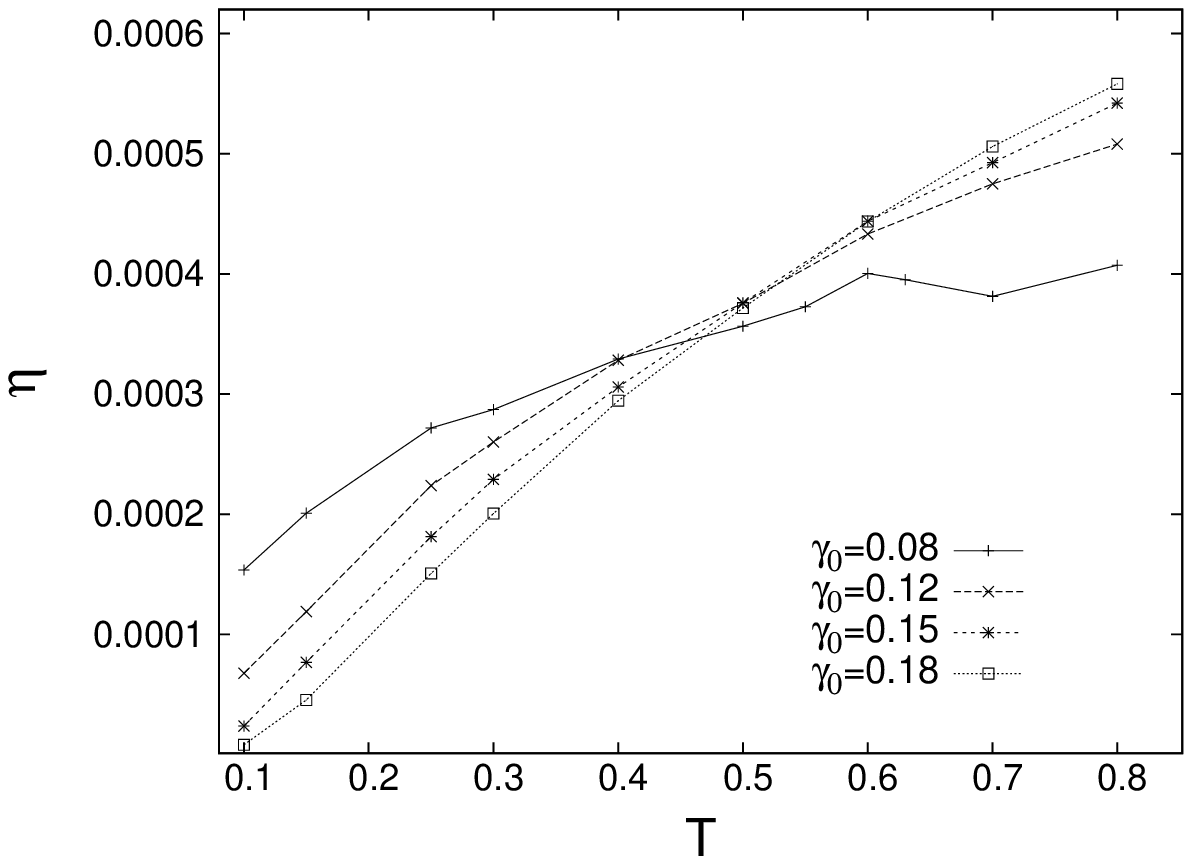}
\vspace*{-5mm}
\caption{The figure shows the variation of $\eta$ as a function of 
$T$ for selected values of various $\gamma_0$ (as indicated in the graph) 
for constant load $L = 0.002$ and $\tau = 8$. Here ${\lambda} = 0.9$, $F_0 = 
0.2$ and $\theta$ = $0.5\pi$.}
\end{figure}

\subsection{Ratchet current against applied load}

We now study the behavior of ratchet current $<\overline{v}>$ by applying a 
load $L$ opposing the current $<\overline{v}>$. Since the numerically 
obtained current $<\overline{v}>$ is negative, we apply $L$ such as to tilt 
the potential with negative slope to oppose the current without load. This 
essentially means that the particle is now subjected to the washboard potential 
  $V(x) = -\sin(x) + xL$, whose average value over 
one period is in the direction of the load. This 
modifies the equation of motion as given below.

\begin{equation}
\frac{d^{2}x}{dt^{2}}=-\gamma(x)\frac{dx}{dt}
-\frac{\partial V(x)}{\partial x} +F(t)+ L +\sqrt{\gamma(x) T}\xi(t),
\end{equation}
with $L>0$. We define the thermodynamic efficiency of the ratchet as
\begin{equation}
\eta=\frac{L<\overline{v}>}{<\overline{W}>/\tau+L<\overline{v}>}.
\end{equation}

Here, we use the natural definition of efficiency as the ratio of work done
by the signal, $F(t)$, against the load $L$ and the total energy spent 
in order to extract that work.

We note that $<\overline{W}>$ is independent of the load $L$ because Eq.
(2.10) is true for each and every trajectory, that is, the mean 
absorbed energy in the presence of load would be the 
same as in the load-free case because $W(0,N\tau)$ depends only on the
explicit time dependence of $U(x,t)$ through $F(t)$. We provide below a brief 
plausible physical explanation on the independence of $\overline{W}$ on $L$, 
to add to the mathematical equality of Eq. (2.10).

The top panel of Fig. 10 shows the effective potential $U(x) = -\sin(x) + xF(t)$ with
$F(t) = F_0\cos(\omega t)$ for the two extreme values of $F(t)$, that is, when $F(t) = F_0$: $U(x) = U_1$,
and when $F(t) = -F_0 $: $U(x) = U_2$. We draw a line corresponding to $U(x) = 0$ for reference.
At the two extreme values of $F(t)$, $U_1$ and $U_2$ are tilted with equal slope but in opposite directions.
If the friction offered by the medium is taken to be uniform, the mean velocities of the particle are equal
in magnitude in both extreme instances of the forcing owing to identical trajectories $x(t)$.
Hence, work done on the system by the external forcing and hence the energy 
dissipated by the system in the two cases are the same. Now let us consider the
situation when, in addition, an external load $L$ is applied to the system. The bottom panel of Fig. 10 shows the potential
profiles ($V_1$ and $V_2$) in the two extreme cases as considered earlier. Clearly, the mean velocities
in the two situations are unequal: one larger than in case of $U_1$, for example, and the other smaller.
This amounts to a mean energy dissipation per cycle of $F(t)$ nearly the same as in the earlier case of
$L=0$.  

In Fig. 11, we plot $<\overline{v}>$ as a function of $L$ for various values of 
$\gamma_0$ for $T=0.1$, $\theta=0.5\pi$, and $\tau=8.0$. We continue to 
increase the load $L$ so long as $<\overline{v}>$ continues to be in the same 
direction as when $L=0$, that is as long as the system does work against the 
load. The variation of $<\overline{v}>$ as a function of $\gamma_0$ at $L=0$ 
for various $\tau$ values are shown to be nonmonotonic in Fig. 8. The 
$<\overline{v}>(L)$ curves appear to be almost straight lines with their slopes 
gradually decreasing with increasing $\gamma_0$ values and their intercepts at
$L=0$ as suggested by Fig. 8. Thus, for large $\gamma_0$ values, for example 
$\gamma_0=0.15$, the curves run almost parallel to the abscissae and hence a 
large load can be applied without $<\overline{v}>$ reversing its direction. On 
the other hand, for smaller values of $\gamma_0$, only a small load can be 
applied.

The changing 'slope' of $<\overline{v}>(L)$ for various $\gamma_0$ have a
another curious fallout too. For example, though at $L=0$, $<\overline{v}>$ 
for $\gamma_0=0.04$ is larger than $<\overline{v}>$ for $\gamma_0=0.08$, as 
the load is increased beyong a certain value, $<\overline{v}>$ for 
$\gamma_0=0.08$ becomes larger than for $\gamma_0=0.04$.  And  this acquired 
inequality continues to remains so over a large range of $L$. It is very 
appropriate to ask, therefore, whether the ratchet becomes more efficient for 
$\gamma_0=0.08$ than for $\gamma_0=0.04$ for the same applied load. And, of
course, whether this trend is true for other $\gamma_0$ values as well as 
suggested by Fig. 11? 

Figure 12 shows the corresponding efficiencies, $\eta$. It shows that $\eta$ is
a nonmonotonic function of $L$ and $\eta$ peaks at larger value of $L$ for
larger $\gamma_0$. At lower loads smaller $\gamma_0$ ratchets are more
efficient than larger $\gamma_0$ ratchets even though the current 
$<\overline{v}>$ is a nonmonotonic function of $\gamma_0$ when $L=0$. However,
for larger $L$, ratchets with larger $\gamma_0$ are more efficient than 
ratchets with smaller $\gamma_0$. This interesting result can, however, be
roughly understood by considering the behaviour of both $<\overline{v}>$ 
and $<\overline{W}>$ as a function of $L$ and $\gamma_0$. Since the product 
$L<\overline{v}>$ is much smaller than $<\overline{W}>$ and $<\overline{W}>$ 
independent of load, $\eta$ is essentially proportional to 
$\frac{L<\overline{v}>}{<\overline{W}>/\tau}$. However, $<\overline{W}>$ is a 
nonmonotonic function of $\gamma_0$ and so is $<\overline{v}>$ which, in 
addition, depends on $L$. In the following we present these results in detail 
separately.

Figure 13 shows $<\overline{v}>$ as a function of $\gamma_0$ for various values 
of load $L$, at $T=0.1$, for $\theta=0.5\pi$ and $\tau=8.0$. It shows that
for a given $L$ below a value $\gamma_0^{min}$ of $\gamma_0$ ratchet does
not do work against the load and hence the concept of thermodynamic efficiency 
is not meaningful. In the inset of Fig. 13, $\gamma_0^{min}$ is plotted as a 
function of $L$.

In Fig. 14 the efficiency $\eta$ is plotted as a function of $\gamma_0$ for
three values of load $L=0.002,~0.003$ and $0.005$ and for other parameters 
same as for Fig. 13. For larger $L$, the curves, which peak at intermediate
$\gamma_0$ values, shift to larger $\gamma_0$ values. From Fig. 14 one can 
extract two important results. For a given load, for all values of $\gamma_0$
smaller than the one at which $\eta$ peaks, ratchets with larger $\gamma_0$
are more efficient than those with lower damping. In these smaller $\gamma_0$
regions for a given $\gamma_0$ efficiency decreases with increasing applied 
load. However, as $\gamma_0$ increases $\eta$ shows nonmonotonic behaviour.
The behaviour includes one in which efficiency increases with applied load
in the large $\gamma_0$ region. This counterintuitive result appears over 
a vast large-$\gamma_0$ region despite the fact that $<\overline{v}>$ 
consistently decreases with increasing load, $L$, Fig. 13, for any fixed
$\gamma_0$. This is because, as mentioned earlier, $\eta$ is essentially
proportional to $L<\overline{v}>$, since $<\overline{W}>$ is independent
of $L$. And $L<\overline{v}>$ increases as $L$ increases since the 
corresponding decrease in $<\overline{v}>$ is smaller in this region of 
$\gamma_0$. This is interesting because more loaded a ratchet is the more 
efficiently it functions!

As has been pointed out earlier\cite{Donrich}, $<\overline{v}>$ increases as a 
function of noise strength (temperature), becomes maximum at arount $T=1.0$
 and then slowly decreases, when all other parameters are kept fixed. In Fig. 
15 we plot $<\overline{v}>$  as a function of $\gamma_0$ for various 
temperatures with $\tau=8.0,~\theta=0.5\pi$ and load $L=0.002$. The curves 
have the same qualitative features as in case of $L=0$, Fig. 8. It can be seen 
that, barring at very small $\gamma_0$ values, $<\overline{v}>$ increases with 
temperature for any value of $\gamma_0$ in the range of temperatures ($T<1.0$) 
we have explored. The corresponding efficiencies are shown in Fig. 16.
 
The efficiency $\eta$ as a function of $T$ at a fixed $\gamma_0$ is given in 
Fig. 17, showing a monotonic increase in the range of $T$ shown. This shows 
that larger efficiencies can be obtained at elevated temperatures. However, 
the increase is likely to stop and even reverse if the temperature is 
increased further. At low temperatures, smaller $\gamma_0$ ratchets are more
efficient than larger $\gamma_0$ ratchets. However, at larger temperatures
larger $\gamma_0$ ratchets become more efficient than the smaller $\gamma_0$
ratchets. Of course, the shown result is for a smaller load of $L=0.002$.
However, the qualitative features remain the same for other loads as well.

\section{Discussion and conclusion}
Admittedly, the frictional ratchet discussed here yields very small currents,
because it is a kind of minimal model having symmetric periodic potential
and driven by a weak subthreshold sinusoidal forcing at a small constant 
temperature. There are ways, as has been reported earlier\cite{comment},
to obtain larger ratchet currents using larger amplitude and smaller frequency
of sinusoidal driving force at elevated temperatures. However, our purpose has 
been to study the ratchet currents in the domain of parameter space where the
probability of obtaining stochastic resonance also is nonzero. Even in this 
restricted domain the thermodynamic efficiency of the ratchet against applied 
load, though small, shows interesting behaviour; efficiency increases with
increasing application of load. This is possible only because the input energy
extracted from the driving field shows nonmonotonic behaviour and has much
larger magnitude than the product of current and load. Also, intuitively one
would feel that as the load is increased the ratchet current should diminish
rapidly. This is indeed true in the small friction range. However, as the 
friction increases the current decreases with increasing load very slowly and
thereby sustains current against much larger loads. Moreover, this system
also allows us to find how the conventional resonance frequency varies with 
increasing damping and large amplitude oscillations. This latter aspect will
be published elsewhere in detail.

We thank the Computer Centre, North-Eastern Hill University, Shillong, for 
providing the High Performance Computing Facility, SULEKOR. 
We are also thankful to the High Performance Computing Laboratory of Computer Science and
Engineering Department, National Institute of Technology
Meghalaya, for the computing facility PARAMSHAVAK.

\end{document}